# KINETIC MODELING OF A SURROGATE DIESEL FUEL APPLIED TO 3D AUTO-IGNITION IN HCCI ENGINES

R. BOUNACEUR[1], P.A. GLAUDE[2], R. FOURNET[3], F. BATTIN-LECLERC[4]

Département de Chimie Physique des Réactions, UMR 7630 CNRS-INPL,
ENSIC, 1, rue Grandville, BP 451, 54001 NANCY Cedex – France
E-mail :
[1]roda.bounaceur@ensic.inpl-nancy.fr
[2]pierre-alexandre.glaude@ensic.inpl-nancy.fr
[3]rene.fournet@ensic.inpl-nancy.fr
[4]Frederique.battin-Leclerc@ensic.inpl-nancy.fr
Tel.: 33 3 83 17 51 25 , Fax : 33 3 83 37 81 20

S. JAY[5], A. PIRES DA CRUZ[6]
IFP, TAE
1 & 4, av. Bois Préau, 92852 Rueil Malmaison Cedex – France
E-mail : [5]Antonio.PIRES-DA-CRUZ@ifp.fr
[6]Stephane.JAY@ifp.fr
Tel.: 33 1 47 52 65 02 , Fax : 33 3 1 47 52 70 68

# KINETIC MODELING OF A SURROGATE DIESEL FUEL APPLIED TO 3D AUTO-IGNITION IN HCCI ENGINES

**INTRODUCTION**

Engines running on HCCI combustion mode (Homogeneous Charge Compression Ignition) have the potential to provide both diesel-like efficiencies and extremely low emissions. HCCI engines rely on a lean combustion process (in excess of air), resulting in a reduction of both particulate matter and nitric oxide (NOx) emissions; the latter are directly related to the creation of tropospheric ozone in urban areas. Fundamentally, the reason for these beneficial effect is that, in a HCCI engine, there is not any region with either fuel rich conditions (that would lead to the generation of particulate matter) or stoichiometric conditions (which would yield high temperature and NOx emissions). These advantages are obtained under optimal working conditions of the engine. In order to warrant such optimal conditions and to further develop HCCI engines, it is important to predict the auto-ignition delay of the homogenous mixture during the compression cycle and to model the formation of pollutants according to the fuel composition (Aceves et al., 2000).

Diesel fuel can potentially be directly used in HCCI engines. Therefore, it is of particular interest to model its auto-ignition characteristics in conditions close to those of diesel engines. However, commercial diesel fuels contain more then 1000 different molecules and their composition is not completely well known. It is then necessary to define a model diesel fuel which would be composed of molecules capable of representing the most important classes of organic compounds present in such a fuel. We have chosen to study a mixture including n-decane as a model molecule of paraffins and α-methylnaphthalene to represent polyaromatic compounds. This hydrocarbon mixture has already been used by other authors (Pitsch and Peters, 1996, Pfahl and Adomeit, 1997, Barths et al., 2000) as a surrogate model for diesel fuel.

The chemistry of hydrocarbon oxidation is very complex and detailed kinetic mechanisms based on elementary steps are a useful way to describe the complex chemical phenomena over the wide range of temperature, pressure and equivalence ratio observed in engines.

The aims of our work were to write a detailed kinetic model for the oxidation and auto-ignition of n-decane and α-methylnaphthalene mixtures, to validate it using laboratory experimental data from the literature and to apply the model to real engine CFD 3D computations. Kinetic mechanism development is described in the first part of this paper. The second part focusses on mechanism validation against experimental laboratory data. The kinetic model is then applied to 3D conventional diesel and HCCI engine simulations. The coupling approach between the detailed chemical mechanism and the engine combustion CFD model is described in the third part of the paper which also presents the CFD simulations and results compared to experimental engine data.

**DETAILED KINETIC MECHANISM FOR THE OXIDATION OF n-DECANE AND α-METHYLNAPHTHALENE MIXTURES**

The mechanism for the oxidation of n-decane/α-methylnaphthalene mixtures includes the mechanism for n-decane automatically generated by EXGAS-ALKANE software as described by Buda et al. (2005)[1], a new mechanism for the oxidation of α-methylnaphthalene and the crossed reactions involved by the interactions between the two fuels. The mechanism for n-decane has been designed to reproduce the low and high temperature oxidation, i.e. the additions of alkyl radicals to oxygen molecules have been taken into account.

---

[1] EXGAS-ALKANES software automatically generates detailed kinetic mechanisms for the oxidation of linear and branched alkanes and is freely available for academic researchers (valérie.warth@ensic.inpl-nancy.fr).

*Mechanism for the oxidation of α-methylnaphthalene*

Few simulations of the oxidation of α-methylnaphthalene are described in the literature: a detailed kinetic mechanism was proposed by Pitsch (1996) and computations of the overall fuel decay in a flow reactor were presented by Shaddix et al. in 1997. The present paper proposes a new mechanism based on the recent models by Da Costa et al. (2003) for the oxidation of benzene and by Bounaceur et al. (2005) for the oxidation of toluene. The α-methylnaphthalene mechanism includes three parts:

♦ The mechanisms for the oxidation of toluene (193 reactions) and benzene (132 reactions) presented in previous papers (Da Costa et al., 2003, Bounaceur et al., 2005) which include the $C_0$-$C_6$ reaction base (709 reactions), first proposed by Fournet et al. in 1999 and updated by Belmekki et al. in 2003. In this reaction base, most $C_0$-$C_2$ and some $C_3$-$C_4$ pressure-dependent rate constants (i.e. unimolecular decompositions, combinations, beta-scissions and additions) follow the formalism proposed by Troe (1974) and efficiency coefficients have been included.

♦ A primary mechanism (Fig. 1) containing 95 reactions, in which only α-methylnaphthalene and oxygen are considered as molecular reactants. The reaction pathways consuming α-methylnaphthalene are: the unimolecular initiations giving H-atoms and phenylbenzyl radicals and naphthyl and methyl radicals, the bimolecular initiations with oxygen molecules abstracting a methyl side hydrogen atom; the ipso-additions (addition of an atom to an aromatic ring leading to the abstraction of another atom or of a radical) of H-atoms producing naphthalene and methyl radicals; the ipso-additions of O-atoms forming methylnaphtoxy radicals and H-atoms; the additions of OH radicals leading to methylnaphthols and H-atoms; the metatheses by H-abstraction. As in the case of toluene, the abstractions of hydrogen atoms were considered from the methyl side, but also from the rings. Pitsch (1996) did not consider the additions of atoms or radicals, neither the abstractions of H-atoms from the rings. The related rate constants are derived from those proposed for toluene (Bounaceur et al., 2005). The reactions of resonance stabilized phenylbenzyl radicals are close to those of toluene: the decomposition by formation of a $C_5$ ring (here, by formation of indenyl radicals and acetylene), the reactions with oxygen molecules to give naphthaldehyde and the combinations with H and O-atoms, OH, $HO_2$ and methyl and phenylbenzyl radicals are taken into account. Products obtained by combination are methyl naphthalene, naphthaldehyde, methylhydroxy naphthalene, methylhydroperoxy naphthalene, which immediately decomposes by breaking of the O-OH bond, ethylnaphthalene and bi-1-methylnaphtyl, respectively. Two differences were nevertheless made in order to obtain satisfactory simulations: the formation of peroxy radicals was not considered and the recombination with $HO_2$ radicals led to the direct formation of naphthaldehyde or naphtyl radicals with the rate constants proposed by Emdee et al. (1992) for a similar reaction in the case of toluene. Naphthyl radicals and the radicals obtained by H-abstractions from the rings react similarly to phenyl radicals, i.e. mainly by reactions with oxygen molecules to give naphthoxy or methylnaphthoxy radicals and by combinations with H and O-atoms, OH, $HO_2$ and methyl radicals. In the case of naphthyl radicals, we have also considered the addition to acetylene to form acenaphthalene as proposed by Pitsch (1996). The reactions of naphthoxy and methylnaphthoxy radicals are directly derived from those of cresoxy radicals, i.e. CO eliminations with rearrangement and terminations with H-atoms. The reactions of resonance stabilized indenyl radicals are inspired from those of cyclopentadienyl radicals: the combinations with H-atoms to give indene or with OH radicals to produce indenol, the self recombinations and the reactions with oxygen atoms or molecules involving the opening of the $C_5$ ring by elimination of CO or $CO_2$.

♦ A secondary mechanism including 152 reactions, in which the reactants are the polycyclic molecular products formed by the primary mechanism, i.e. indene, methylindene, naphthalene, dimethylnaphthalenes, α-ethylnaphthalene, acenaphthalene, bi-indenyl, bi-1-methylnaphtyl, indenol, naphthols, α-naphthaldehyde, methylnaphthols, α-methylhydroxy naphthalene, naphthalenediones and methylnaphthalenediones (see fig. 1). The different involved isomers are not differentiated. The secondary mechanism of a polycyclic molecule has been written based on the reactions of the analogous monocyclic parent proposed for modelling the oxidation of benzene (Da Costa et al., 2003) and toluene (Bounaceur et al., 2005). For instance, the reactions of naphthalene are derived from those of benzene and those of naphthaldehyde from benzaldehyde and both compounds lead mainly to the formation of naphthyl radicals.

Specific heat, heat of formation and entropy of the molecules or radicals have been calculated using software THERGAS (Muller et al., 1995), based on the group and bond additivity methods proposed by Benson in 1976. As simulations showed an important sensitivity to thermodynamic data, the heats of formation of some biaromatic species (naphthalene, α-methylnaphthalene, naphthols (taken as that of 1-naphthol), indene, acenaphthalene and indenyl, naphthyl, phenylbenzyl and naphthoxy radicals) for which the presence of two aromatic cycles increases the uncertainty in the group additivity estimation, have been taken from the values computed by Richter et al. (2000) using density functional theory.

*Crossed reactions in n-decane/α-methylnaphthalene mixtures*

While previous studies have shown that crossed reactions have negligible effect on the oxidation of mixtures of two alkanes (Glaude et al., 2002) or even of an alkane and toluene (Klotz et al., 1998), we have considered here the crossed reaction which can be involved by the presence of important amounts of resonance stabilized phenylbenzyl radicals and which could enhance the inhibiting effect of α-methylnaphthalene only induced by an action on the common pool of free radicals. Two types of possible crossed reactions have been taken into account :

- Metatheses involving the abstraction by a phenylbenzyl radical of a hydrogen atom from n-decane. Phenylbenzyl radicals can be involved in the following reversible metatheses with n-decane and lead to the formation of the 5 isomers of decyl radicals:

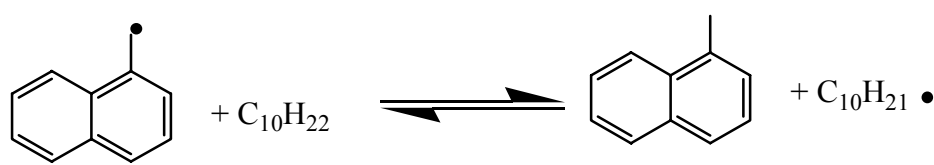

- The combination of phenylbenzyl and decyl radicals. Phenylbenzyl radicals can also combine with each of the 5 isomers of decyl radicals:

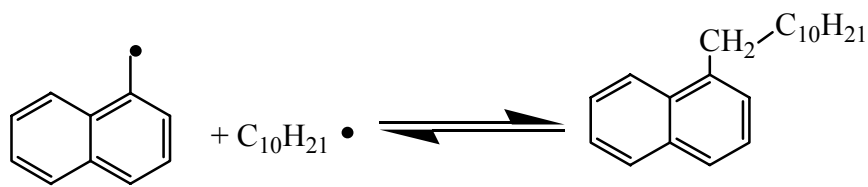

Similar types of crossed reactions have been considered for mixtures containing n-butane and toluene and have allowed autoignition delay times obtained in a shock tube to be modelled satisfactorily.

**VALIDATION OF THE DETAILED KINETIC MECHANISM USING LABORATORY EXPERIMENTS**

The mechanism for the oxidation of pure α-methylnaphthalene has first been validated and only then simulations have been performed for mixtures of α-methylnaphthalene and n-decane. The global mixture mechanism involves 2834 reactions and 530 species among which 1282 reactions and 189 species belong to the α-methylnaphthalene mechanism. Simulations have been performed using software components from the CHEMKIN II library (Kee et al., 1993). In all graphical representations in this section, symbols correspond to experimental data and lines represent the numerical simulations.

*Validation of the α-methylnaphthalene oxidation mechanism*

The oxidation of α-methylnaphthalene has been experimentally investigated in several types of reactors in a temperature range varying from 800 K to 1500 K and a pressure range from 1 bar to 13 bar.

Pfahl et al. (1996) have measured auto-ignition delay times for α-methylnaphthalene in a shock tube at temperatures from 900 K to 1500 K, a pressure of 13 bar and a stoichiometric mixture in air (1.6 % hydrocarbon). Figure 2 shows that our simulations reproduce well these experimental data.

Shaddix et al. (1992, 1997) have studied the oxidation of α-methylnaphthalene in a flow reactor at 1170 K, atmospheric pressure, for mixtures containing around 0.1% hydrocarbon with nitrogen as bath gas. Globally, there is an acceptable agreement between our calculations and the measurements for the three studied equivalence ratios, 0.6, 1 and 1.5: The consumption of α-methylnaphthalene (Fig. 3a) is very well reproduced by the model; the formation of naphthalene (Fig. 3b) is also well reproduced although the agreement is not as good as for the α-methylnaphthalene; the consumption of oxygen and the production of carbon monoxide, naphthaldehyde and naphthols (Fig. 3c) are in agreement with experimental measurements for an equivalence ratio of 0.6.

Marchal (1997) has studied the oxidation of α-methylnaphthalene in a jet stirred reactor between 800 K and 1150 K, at 10 bar, a residence time of 0.5 s, for a stoichiometric mixture containing 0.111% hydrocarbon with nitrogen as bath gas. The agreement is here also acceptable, as shown in Fig. 4 for the consumption of α-methylnaphthalene and oxygen and the formation of carbon monoxide and carbon dioxide (Fig. 4a), methane, benzene, toluene and naphthalene (Fig. 4b). The orders of magnitude for product formation are well reproduced, but the maxima are obtained for temperatures around 40 K higher than in experiments.

*Validations of the mechanism for the oxidation of n-decane/α-methylnaphthalene mixtures*

No gas-phase study of the oxidation n-decane/α-methylnaphthalene mixtures was found in the literature, but Moriue et al. (2000) have studied the auto-ignition of these mixtures in a pressurized chamber at 3 bar in which a suspended droplet was suddenly brought to a temperature from 600 K to 750 K. Diagnostics carried out with a Michelson interferometer detected cool-flame, as well as hot flame appearance. The purpose of the current validation is not to propose an accurate physical model of these experiments, but to roughly reproduce the qualitative trends observed by using our detailed chemical mechanism. We have then simulated this system as an adiabatic well-mixed reactor containing, after an induction time due to vaporization ($\tau_{vap}(T)$), a gas phase stoichiometric fuel/air mixture. In the real reactor, there is a concentration gradient which develops around the droplet when it vaporizes that is not reproduced by simulations. Nevertheless, it can be assumed that the appearance of cool flame and hot ignition will be triggered by the most reactive zone in which the composition is probably not far from stoichiometry. For each studied temperature (T), the vaporization induction time is deduced from a comparison between the experimental ignition delay time for cool flame appearance ($\tau_{cf\ exp.}(T)$) and the time simulated assuming gas-phase conditions ($\tau_{cf.\ sim\ gas\text{-}phase}(T)$) in the case of pure n-decane:

$$\tau_{vap}(T) = \tau_{cf\ exp.}(T) - \tau_{cf\ sim\ gas\text{-}phase}(T).$$

Since the pure n-decane model has been validated independently for real gas-phase experiments

(Buda et al., 2005), $\tau_{\text{cf. sim gas-phase}}(T)$ can be taken as a baseline. For n-decane/α-methylnaphthalene mixtures, the same $\tau_{\text{vap}}(T)$ is used for all compositions and the simulated ($\tau_{\text{cf sim}}(T)$) ignition delay time for cool flame appearance is then :

$$\tau_{\text{cf sim}}(T) = \tau_{\text{vap}}(T) + \tau_{\text{cf sim gas-phase}}(T).$$

Fig. 5a shows a comparison between experimental and simulated ignition delay times for cool flame appearance for mixtures containing different fractions of α-methylnaphthalene. In spite of the strong assumptions made here, the agreement in terms of auto-ignition delays as a function of temperature is good for the mixture containing 40% α-methylnaphthalene, but deteriorates with higher contents of this component, for which the reactivity is overestimated by simulations. This is probably due to the difference in volatility between n-decane (normal boiling point = 447.3 K) and α-methylnaphthalene (normal boiling point = 517.9 K), inducing a higher content of n-decane in the gas-phase than in the liquid phase. Fig. 5b shows a comparison between experiments and simulations for the time between cool flame and hot-flame appearance. Our model qualitatively reproduces the inhibiting effect of the addition of α-methylnaphthalene, but the simulated effect is more marked for mixtures containing from 0 to 40 % aromatic compounds than in experiments. This is also probably due to the difference in volatility between both hydrocarbons not taken into account by the numerical simulations. These validations should be regarded as a very primary approximation showing that the crossed reactions considered are not meaningless, but further work is needed when new experimental data in a gas-phase reactor is available.

# APPLICATION OF THE DETAILED KINETIC MECHANISM TO ENGINE 3D SIMULATIONS

*Delay sensitivity analysis in engine conditions*

In the absence of more realistic diesel fuel models, n-heptane has been widely used in past studies as a surrogate diesel fuel (Colin et al., 2005, Curran et al., 1998-1, Curran et al., 1998-2, Embouazza et al., 2003 and Pires da Cruz, 2004). Detailed kinetic models for the auto-ignition and oxidation of n-heptane have been available for a number of years. It is therefore interesting to compare the auto-ignition behavior of n-heptane against n-decane and α-methylnaphthalene mixtures in engine thermodynamic conditions. In this study, we have first compared the relative deviations between the main ignition delays for pure n-heptane and a mixture of 70 % volumetric n-decane and 30 % α-methylnaphthalene (hereafter named *Z*=70 mixture, where *Z* is the percentage of α-methylnaphthalene in the mixture). According to Pitsch and Peters (1996), the 70/30 proportion is the one which best represents the behavior of the diesel fuel. The n-heptane mechanism was generated by the same algorithm as the one for n-decane. Both are issued from the EXGAS software.

Auto-ignition delays with both mechanisms can be calculated in an adiabatic, homogeneous reactor at constant pressure along the thermodynamic conditions that one may expect to find in a real engine. The differences between the delays of these two fuels can then be represented by plotting the relative deviations between the two mechanisms in an initial temperature versus initial pressure $(T_0,p_0)$ space for different compositions distributed along all possible conditions that we have tested. The initial temperature $T_0$ and the initial pressure $p_0$ were varied ($T_0 \in$ [600 K-1500 K] with a 10 K step and $p_0 \in$ {1,1.5,2,2.5,3,4,6,8 MPa} respectively). The mixture composition expressed in terms of the global equivalence ratio $\phi$ ($\phi \in [0.3$-$0.5$-$0.7$-$1.0$-$1.5$-$3.0]$) and dilution gas fraction relative to air composed of 21 % $O_2$ and 79 % $N_2$ ($X_{res} \in [0\%$-$30\%$-$60\%$-$80\%$-$90\%]$) was also varied.

The initial molar composition for the detailed kinetics computations ($X_{fuel}$, $X_{O_2}$, $X_D$, where $X_D$ is the molar fraction of dilution gases) was determined as follows :

$$\phi = \alpha \frac{X_{Fu}}{X_{O_2}} \text{ with } \alpha = 11 \text{ for } nc_7h_{16}$$

$$X_{res} = 1 - \frac{4.76 \; X_{O_2}}{X_{O_2} + X_D}$$

$$X_{Fu} + X_{O_2} + X_D = 1$$

One can note that, in this system, $X_{res}$ is seen as the amount of non-reactive species ($N_2$, $CO_2$, $H_2O$, Ar,…) in excess relative to the amount of $N_2$ present in air. In some cases, one needs to be careful that the definition adopted here matches the definition of $X_{res}$ used for experimental results.

An example of the results obtained is plotted in Fig. 6a for $\phi = 1$ and $X_{res} = 0\%$.

The differences essentially depend on the initial temperature. Pressure effects are small. Nevertheless, the same deviation between two distinct pressures will result in a more important difference between the delays when the pressure is low (for instance, p < 3 MPa) since the auto-ignition delay is longer in such conditions. The discrepancies between the two fuels are pressure independent and are essentially located in a medium temperature range, between 750 K and 900 K (cool flame region). Deviations between delays for both fuels can be significant and reach 100 % for these conditions. Above 1000 K, differences fall below 10 %. These results are confirmed by the classical delay plots as a function of pressure and inverse of the temperature (see Fig. 6b). One may then expect that the good prediction of the auto-ignition delays like those obtained by Réveille et al. (2004) in engine 3D simulations with pure n-heptane at high temperature will not be affected. When low temperature mechanisms are involved, some strong effects can be observed by choosing the n-decane/α-methylnaphthalene mixture.

*Coupling approach with the 3D combustion model ECFM3Z of Colin and Benkenida (2004)*

In practical applications, especially for new low emission combustion concepts such as HCCI engines, the cool flame heat release $C_1\Delta h$, where $C_1$ is the fuel fraction consumed by the cool flame and $\Delta h$ is the enthalpy difference due to combustion, significantly contributes to the total heat release. Such strategies imply that the mixture is nearly homogeneous at an early time in the engine cycle so that its temperature is not high enough to trigger fast auto-ignition. Bypassing the cool flame mechanism by considering a single delay $\tau_{HT}$ without the early heat release $C_1\Delta h$ may imply an erroneous calculation of the main auto-ignition. As shown by Pires da Cruz (2004), the heat released during the cool flame period (during $t_{LT} < \tau < t_{HT}$ in Fig. 7) significantly modifies the thermodynamic conditions (pressure and temperature increase). Consequently, the chemical reactions after the delay $\tau_{LT}$ must be described accurately.

The TKI (Tabulated Kinetic Ignition) auto-ignition model of Colin et al. (2005) is briefly summarized here. It relies on tabulated auto-ignition quantities issued from detailed chemistry calculations. The auto-ignition database can be generated by the Senkin code (Lutz et al., 1987) which is part of the Chemkin package (Kee et al., 1989). Databases have been built for the same set of thermodynamic conditions ($T_0, p_0, \phi_0, X_{res_0}$) as used in Pires da Cruz (2004) at constant pressure and using the model fuels described in this work (n-heptane and the Z=70 mixture).

The following parameters are determined through post-processing of the complex chemistry simulations (see Fig. 7) and then tabulated in the databases:

- Cool flame ignition delay $\tau_{LT}$ when present. It corresponds to the maximum temperature versus time derivative during the cool flame ignition period.

- Cool flame heat release $C_1$. Even if in some cases the cool flame heat release is not fast, it is always modeled as a step. If a cool flame is not present, $C_1 = 0$ and the main auto-ignition process is directly computed using chemical progress rates $\dot{\omega}_c$, where $c$ is a progress variable between 0 and 1 based on the temperature evolution.

- Chemical progress rate $\dot{\omega}_c$. The time derivative of the progress variable $c$ is stored for discrete values of this quantity. An adequate refinement is needed in the low temperature region. This corresponds to small values of $c$. Different number and combinations of these points have been tested. Sufficient accuracy was obtained with 7 points between $c = 0.025$ and $c = 0.3$. For $c > 0.3$, the reaction is always very fast and $\dot{\omega}_c$ becomes sufficiently high to promote auto-ignition.

The TKI auto-ignition model was implemented in the IFP-C3D engine code. Auto-ignition information is retrieved from the databases through linear interpolation along the local thermodynamic conditions met in a computational cell. This model allows computation of the auto-ignition processes in both low and high temperature conditions. It was validated against Senkin detailed chemistry calculations performed on a constant volume configuration and on a perfectly homogeneous rapid compression machine system. The results presented on Fig. 8 show that the TKI model is capable of reproducing the auto-ignition behavior of the Z=70 mixture when compared to detailed chemistry calculations for conditions typical of HCCI engines ($\phi$=0.7, EGR=0 and 30%, engine speed of 1200 rpm…).

*Engine 3D simulations*

Engine computations with the IFP-C3D code were performed using thermodynamic properties of a diesel fuel and kinetic results of either pure n-heptane or the mixture Z=70. The computed cases are 3D sector geometries operating on conditions shown in Table 1. 3D temperature fields corresponding to both conditions on Table 1 are plotted in Fig. 9. One can see that the onset of auto-ignition is different between the two cases. In Fig. 9a, the conventional part load temperature field shows that auto-ignition occurs in a narrow region. Reaction rates in this case are high. In Fig. 9b, the HCCI engine computation features a smoother heat release and the auto-ignition region spreads over a wide area with cooler conditions. The temperature fields in Fig. 9 clearly show the differences between the conventional diesel engine and the HCCI operating conditions in terms of the homogeneity of the mixture at the onset of auto-ignition. The lower temperature field in HCCI conditions strongly justify the particular attention dedicated to cool flames description included in the surrogate fuel kinetic modeling and in the engine auto-ignition 3D model exposed previously.

Figure 10a compares the experimental pressure curve of a partial load, diluted (EGR=30%), direct injection diesel engine with the pressure curves computed with n-heptane and with the Z=70 mixture fuel. Only small differences were observed on the auto-ignition and remaining combustion processes between experimental and modeling results using both model fuels.

Figure 10b corresponds to a HCCI (EGR = 50%) split injection engine. Again, in these conditions, the n-heptane and the Z=70 mixture mechanisms predict the same auto-ignition delay. However, the pressure rise after ignition is slower with the Z=70 model indicating a lower heat release rate when n-decane and α-methylnaphthalene kinetic properties are considered. It is important to note that the mixing process plays a significant role in the auto-ignition prediction which may explain why the initial delay and cool flame heat release is not so well positioned for both n-heptane and n-decane/α-methylnaphthalene kinetics. It can be deduced from these results that the simulations are more sensitive to the kinetics in HCCI conditions where the auto-ignition delays are usually longer when compared to conventional diesel. The use of detailed kinetics is essential in those conditions insofar as large deviations between delays and reaction rates are observed between the two fuels.

**CONCLUSION**

A new detailed kinetic mechanism has been proposed to model the oxidation of α-methylnaphthalene and n-decane/α-methylnaphthalene mixtures. The mechanism for α-methylnaphthalene is derived from recent mechanisms for the oxidation of benzene and toluene and has been successfully validated using experimental results obtained in a shock tube, in a flow reactor and in a jet-stirred reactor. The mechanism for the mixture can qualitatively reproduce the evolution of the cool flame and auto-ignition times observed when introducing a fuel droplet in a heated pressurized chamber.

The n-decane/α-methylnaphthalene mixture kinetic model developed in this study as well as an available n-heptane auto-ignition mechanism were applied to both conventional diesel and HCCI engine 3D simulations. The kinetic mechanisms were not directly coupled with the 3D CFD engine code but rather used a priori in the construction of auto-ignition delays and reaction rate databases covering a broad range of thermodynamic conditions typical of internal combustion engines. The engine simulation results show that both fuels behave well in predicting auto-ignition delays and reproducing the experimental pressure throughout the engine cycles. This explains why, from a strict auto-ignition delay viewpoint, both diesel surrogates (n-heptane and the Z=70 mixture) can be used as model fuels. However, in HCCI engine conditions, important differences were observed in the reaction rate after ignition. Since pollutant emissions from engines are mostly generated during the high temperature combustion phase, it is important during this period to have a correct estimation of the reaction rate. Therefore, a surrogate diesel fuel with kinetic properties closer to a real diesel (including alkanes, aromatics and eventually other components) should perform better than a single component fuel such as n-heptane.

**FIGURE CAPTIONS**

Figure 1: Primary mechanism of the oxidation of α-methylnaphthalene. The bicyclic primary products are written in bold.

Figure 2: Auto-ignition of α-methylnaphthalene in a shock tube (Pfahl et al., 1996).

Figure 3: Oxidation of α-methylnaphthalene in a flow reactor (Shaddix et al., 1992, 1997): (a) mole fraction of α-methylnaphthalene, (b) mole fraction of naphthalene and (c) mole fraction of oxygenated compounds for an equivalence ratio of 0.6. The initial hydrocarbon mole fractions used for simulations were 1100 ppm at Φ 0.6 (a time shift of 20 ms has been considered) and 950 ppm at Φ = 1 and Φ = 1.5 (no time shift, but $T_{sim}$ = 1190 K).

Figure 4: Oxidation of α-methylnaphthalene in a jet-stirred reactor (Marchal, 1997).

Figure 5: Auto-ignition of n-decane/α-methylnaphthalene mixtures in a pressurized chamber with liquid injection (Moriue et al., 2000): (a) ignition delay times for cool flame appearance ($\tau_{cf}$) and (b) time between cool flame appearance and hot flame appearance ($\tau_{igni}$ - $\tau_{cf}$).

Figure 6: Comparisons of the n-heptane and the n-decane/ α-methylnaphthalene mixture delays as a function of 1000/T. (a) Relative deviations (in %) between the delays of the two fuels, (b) pure n-heptane delay curves (symbols) and n-decane/ α-methylnaphthalene delay curves (solid lines).

Figure 7: Coupling approach between detailed kinetics and 3D combustion model. Right : Model schematic using a double tabulated delay ($\tau_{LT}$, $\tau_{HT}$) and cool flame consumption $C_1$. Left : Tabulated database parameters ($\tau_{LT}$, $C_1$, reaction rates as a function of the progress variable c) for the TKI 3D auto-ignition model used in this paper.

Figure 8: Comparison of temperature profiles obtained with the TKI 3D auto-ignition model (long dashed line) and with detailed chemistry computations (solid line) for the Z=70 mixture. $\phi = 0.7$, $X_{res} = 0$ % and 30%. Engine speed is 1200 rpm and at $ca = -180$, $T_0 = 550$ K, $p_0 = 0.4$ MPa.

Figure 9: Engine 3D simulations. Temperature field (sector mesh) in the combustion chamber at the onset of auto-ignition. (a) Low load direct injection diesel engine (EGR = 31 %, BMEP = 3.7 bar, 1640 rpm), (b) split injection HCCI engine (EGR = 50 %, BMEP = 5bar, 1500 rpm).

Figure 10: Engine 3D simulations. Comparison of the pressure evolutions using n-heptane and n-decane/α-methylnaphthalene. (a) Low load direct injection diesel engine (EGR = 31 %, BMEP = 3.7 bar, 1640 rpm), (b) split injection HCCI engine (EGR = 50 %, BMEP = 5 bar, 1500 rpm).

**Table 1: Engine test cases: 3D simulations main parameters.**

| Simulation parameters | Part load (1 injection) (related to fig. 9a) | HCCI (split injection) (related to fig. 9b) |
|---|---|---|
| BMEP[2] | 0.37 MPa | 0.5 MPa |
| RPM[3] | 1640 | 1500 |
| EGR[4] rate | 31 % | 50 % |
| Compression ratio | 17.8 | 14 |
| Injection start | 6 ca BTDC[5] | 9 ca and −11 ca BTDC |
| Injection duration | 8 ca | 3.79 ca and 3.54 ca |
| Injection pressure | 96 MPa | 140 MPa |

---

[2] Brake Mean Effective Pressure

[3] Rotations Per Minute

[4] Exhaust Gas Recirculation

[5] Before Top Dead Center

FIGURE 1

FIGURE 2

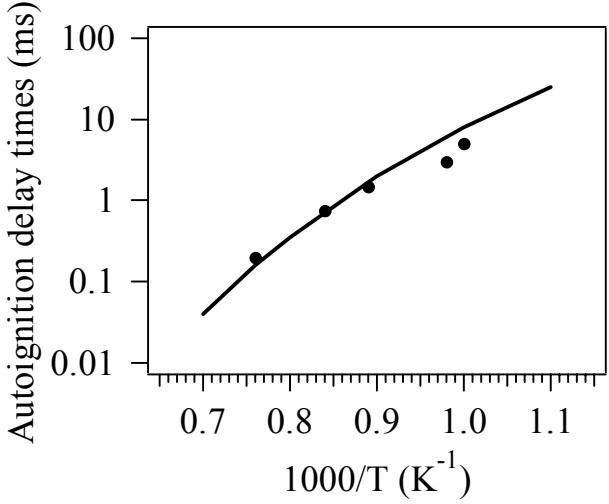

FIGURE 3

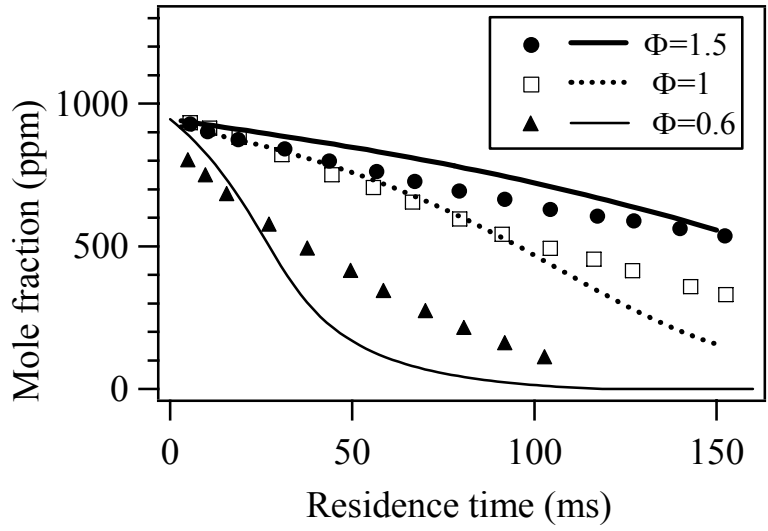

(a)

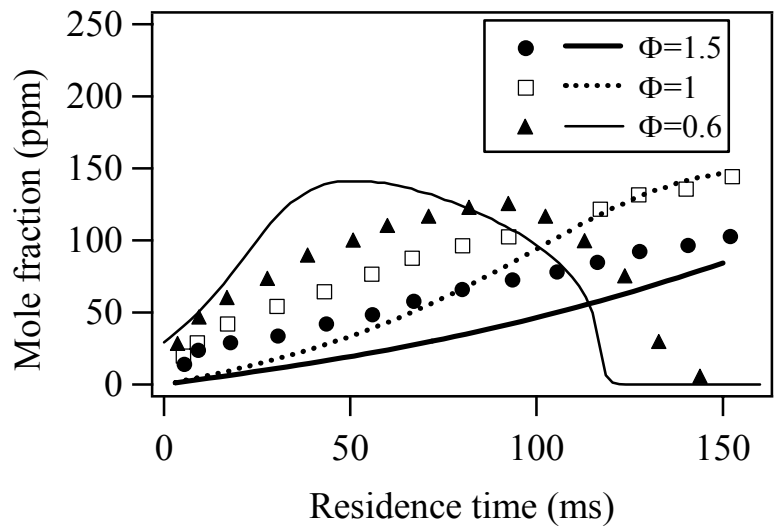

(b)

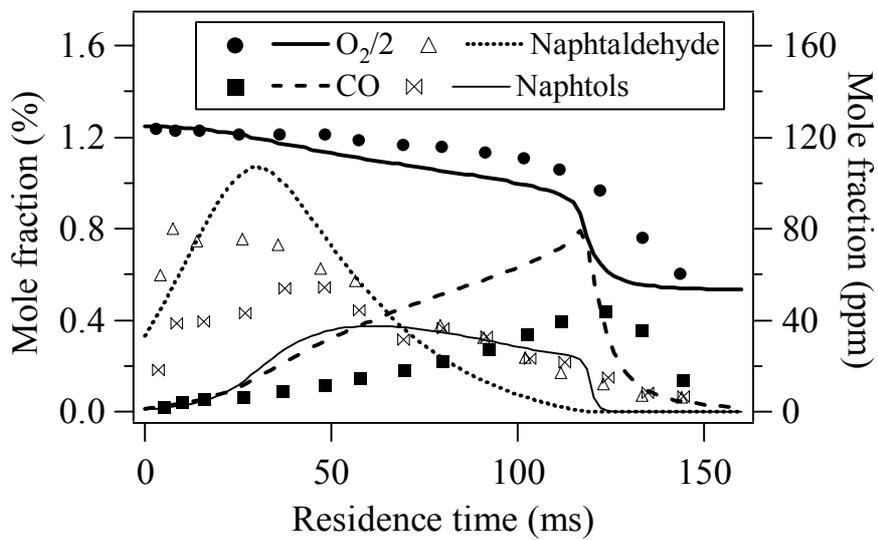

(c)

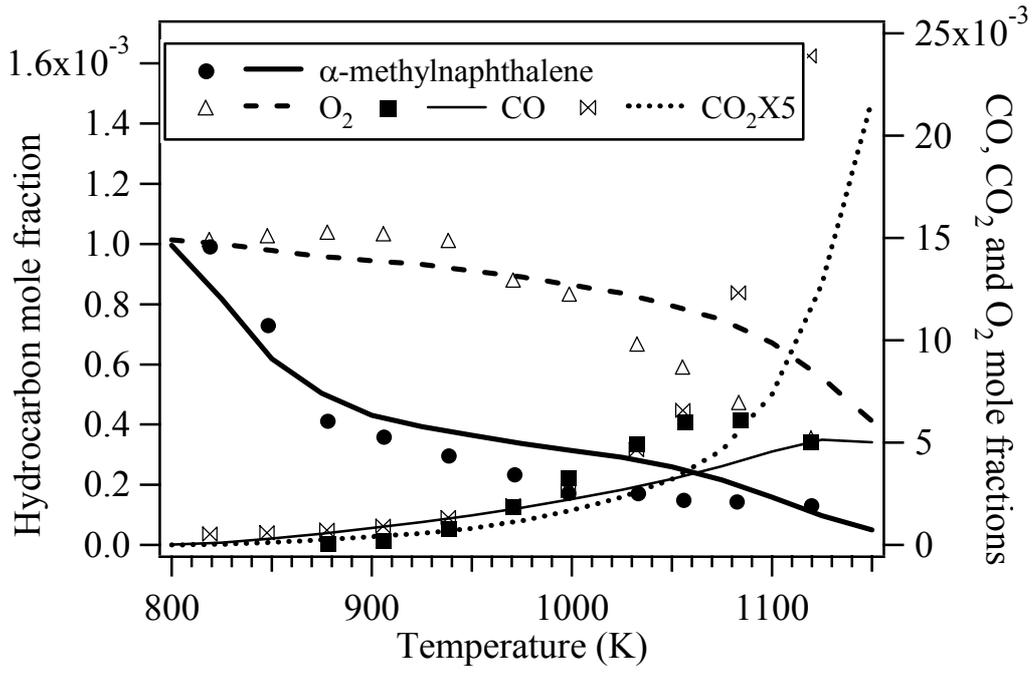

(a)

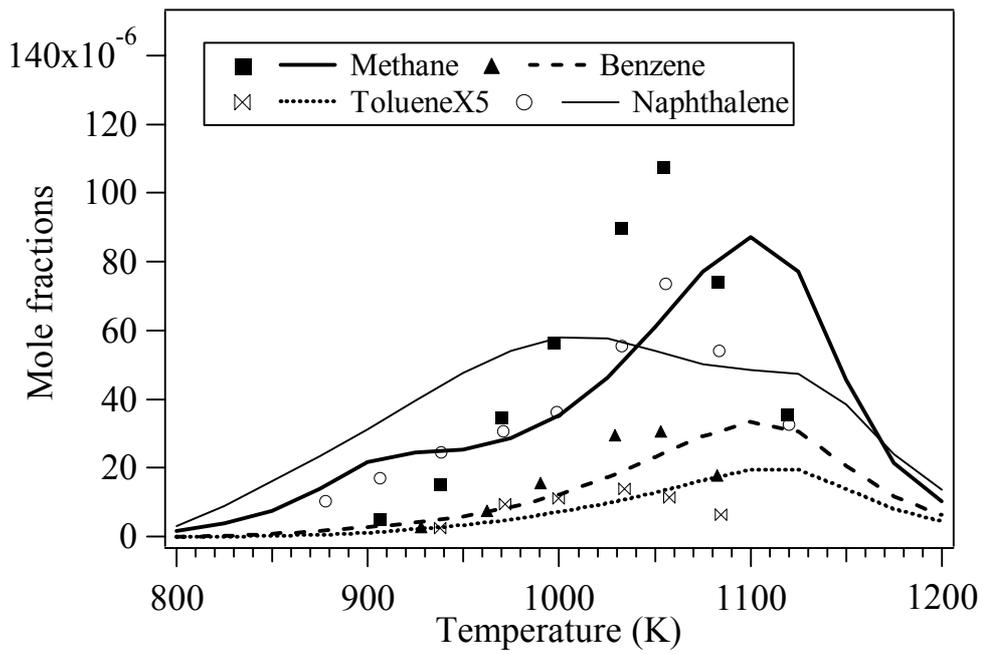

(b)

FIGURE 5

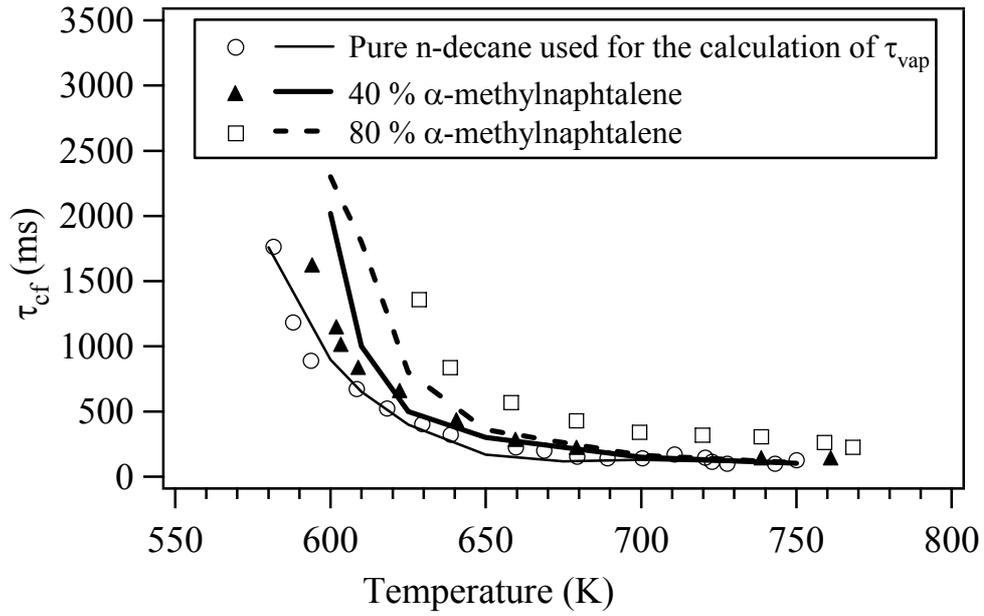

(a)

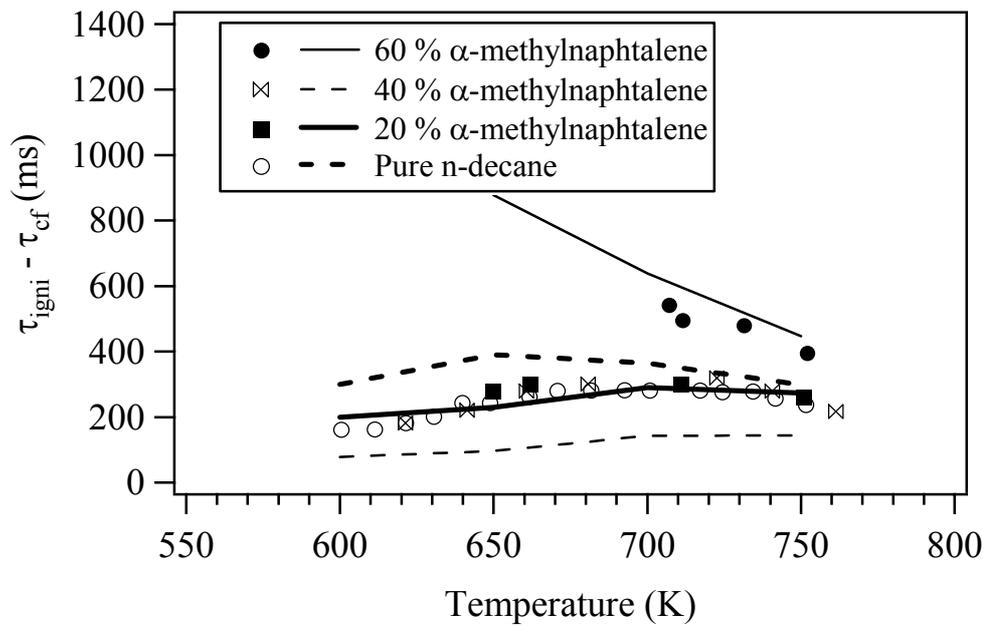

(b)

FIGURE 6

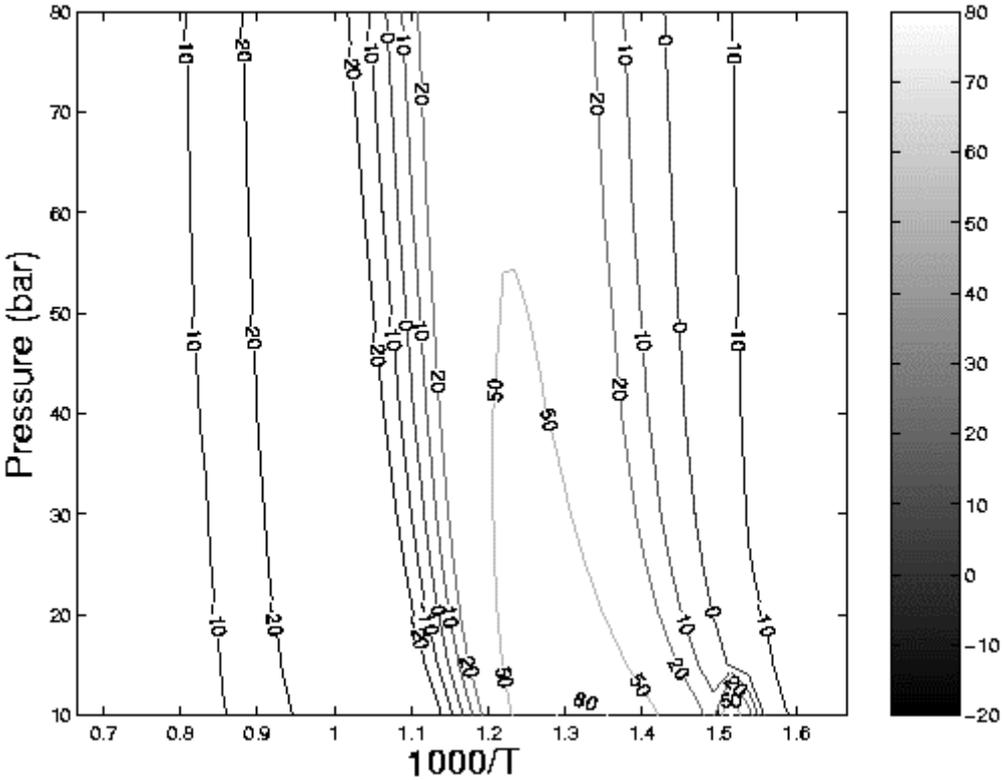

(a)

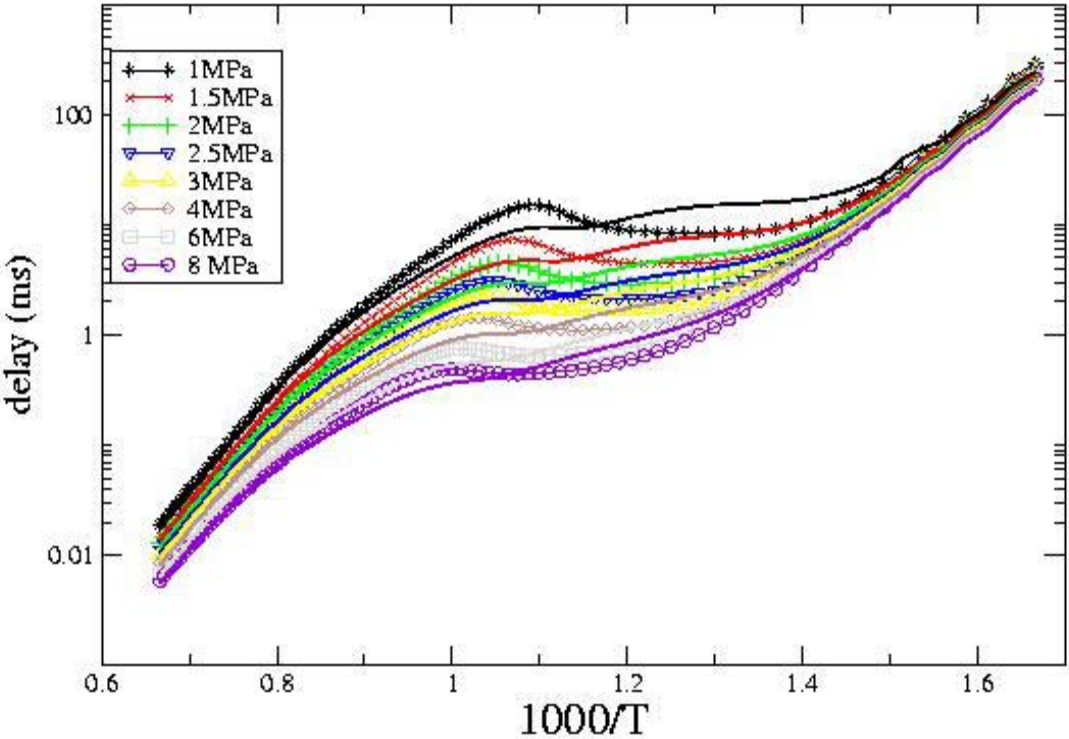

(b)

FIGURE 7

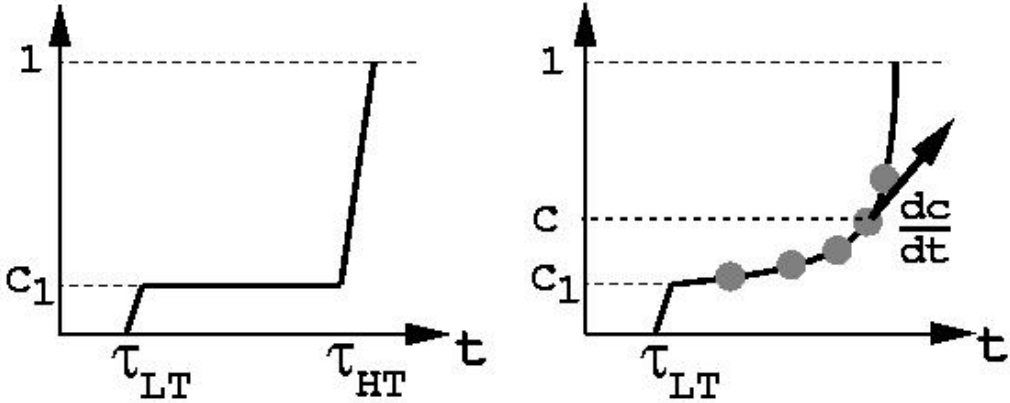

FIGURE 8

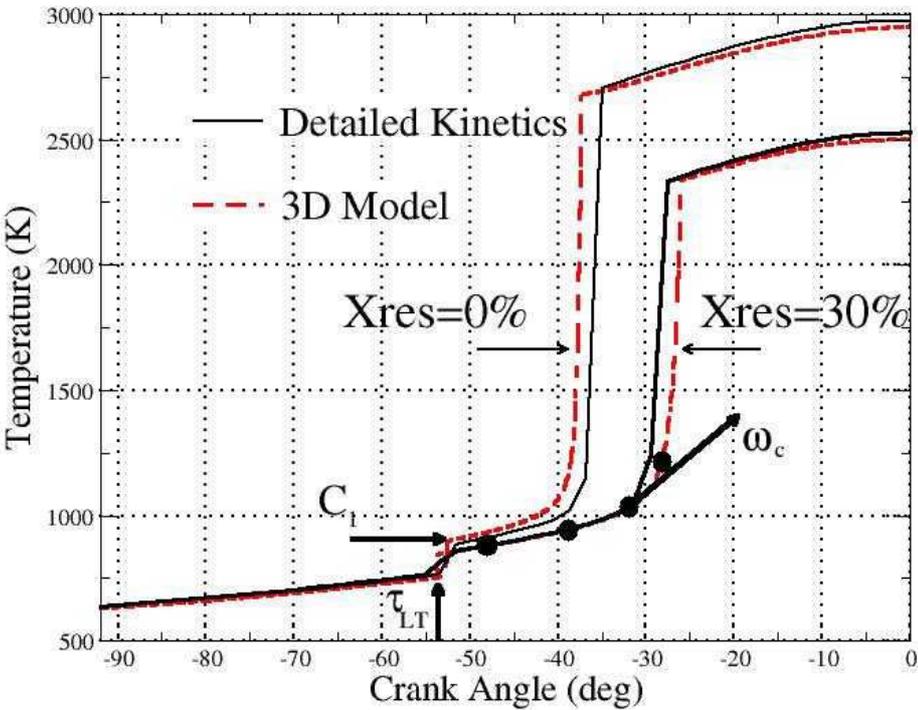

FIGURE 9

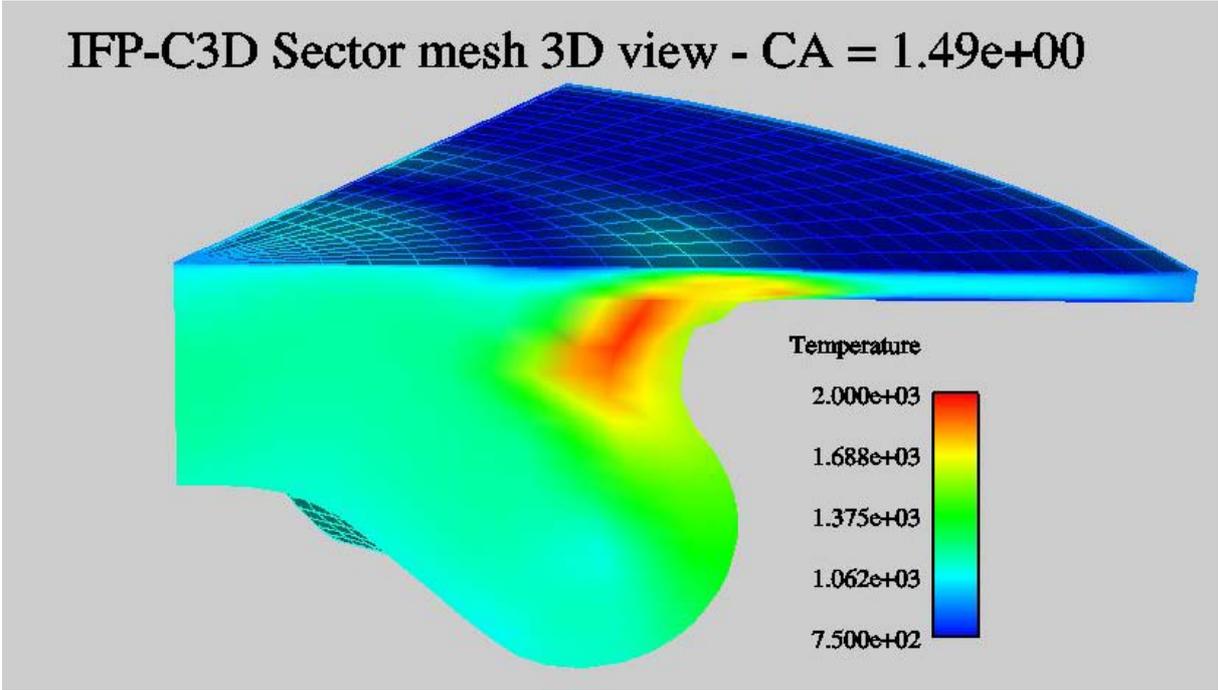
(a)

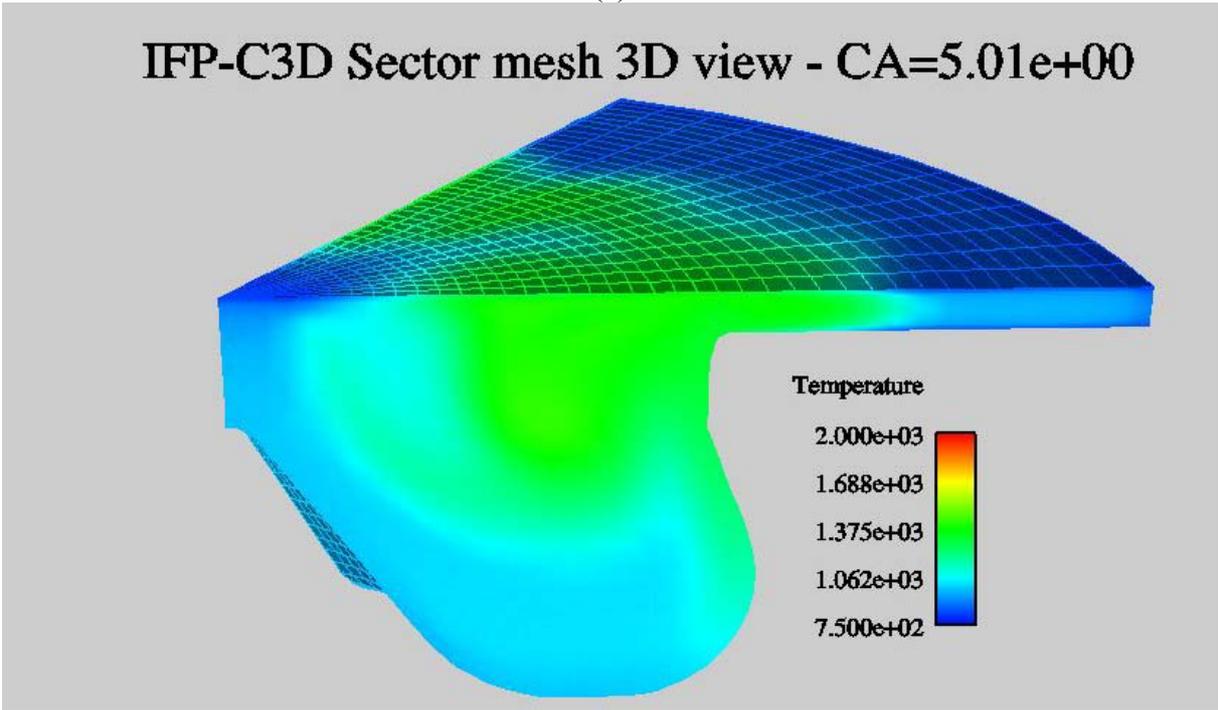
(b)

FIGURE 10

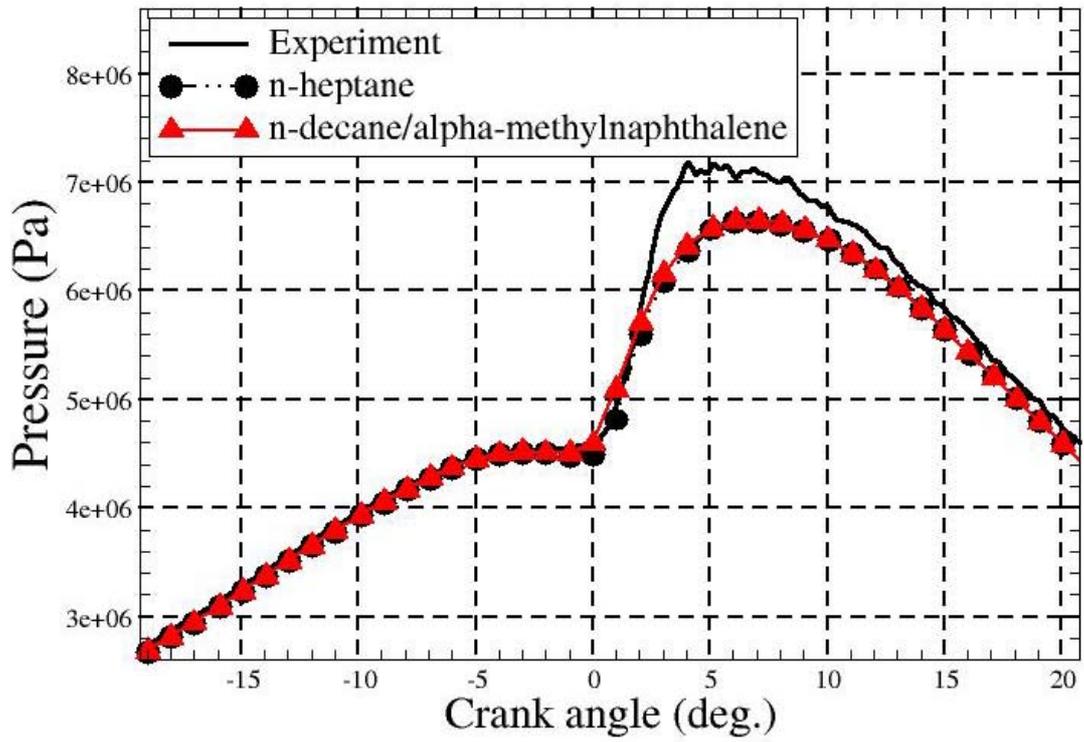

(a)

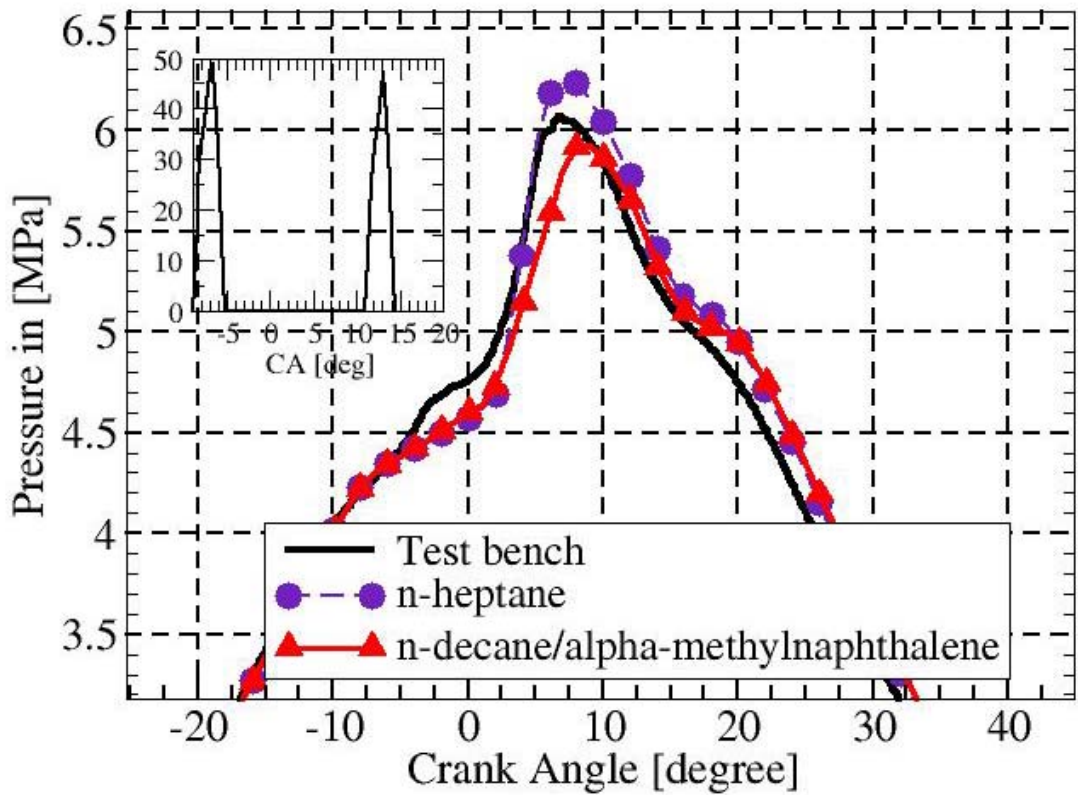

(b)

**ABSTRACT**

The prediction of auto-ignition delay times in HCCI engines has risen interest on detailed chemical models. This paper described a validated kinetic mechanism for the oxidation of a model Diesel fuel (n-decane and α-methylnaphthalene). The 3D model for the description of low and high temperature auto-ignition in engines is presented. The behavior of the model fuel is compared with that of n-heptane. Simulations show that the 3D model coupled with the kinetic mechanism can reproduce experimental HCCI and Diesel engine results and that the correct modeling of auto-ignition in the cool flame region is essential in HCCI conditions.

Keywords : Diesel fuels, n-decane, α-methylnaphthalene, auto-ignition, HCCI engines.

**Authors :**

Roda Bounaceur : PhD in chemical engineering (2001), CNRS research engineer since 2004.

Pierre-Alexandre Glaude : PhD in chemical engineering (1999), CNRS researcher since 2000.

René Fournet : PhD in Physical Chemistry (1994), habilitated (2002), lecturer at ENSIC-INPL since 1995

Frédérique Battin-Leclerc : PhD in chemical engineering (1991), habilitated (1997), CNRS researcher since 1991.

Antonio Pires da Cruz: PhD in Mechanical Engineering (1997); Post-doctoral at ExxonMobil (1998-2000); IFP Research Engineer since 2000; IFP-School assistant Professor since 2005.

Stéphane Jay: PhD in Mechanical Engineering (2003); IFP Research Engineer since 2003.